# Ageing memory and glassiness of a driven vortex system.

Xu Du[1], Guohong Li[1], Eva Y. Andrei[1], M. Greenblatt[2], P. Shuk[2]

[1]*Department of Physics and Astronomy, Rutgers University, Piscataway, New Jersey 08855 USA*

[2]*Department of Chemistry, Rutgers University, Piscataway, New Jersey 08855, USA*

**Many systems in nature - glasses[1-11], interfaces[12] and fractures[13] being some examples - cannot equilibrate with the environment, which gives rise to novel and surprising behaviour such as memory effects, ageing and nonlinear-dynamics. Unlike their equilibrated counterparts, the dynamics of out-of-equilibrium systems is generally too complex to be captured by simple macroscopic laws[1]. Here we investigate a system that straddles the boundary between glass and crystal: a Bragg glass[14,15], formed by vortices in a superconductor. We find that the response to an applied force evolves according to a stretched exponential, with the exponent reflecting the deviation from equilibrium. After the force is removed, the system ages with time and its subsequent response time scales linearly with its "age" (simple ageing), meaning that older systems are slower than younger ones. We show that simple ageing can occur naturally in the presence of sufficient quenched disorder. Moreover, the hierarchical distribution of time scales, arising when chunks of loose vortices cannot move before trapped ones become dislodged, leads to a stretched exponential response.**

Glassy states of matter abound with seeming contradictions: macroscopically they are rigid like crystals but microscopically their structure is closer to that of liquids. At the same time, their response to external drives is unlike either that of crystals or liquids displaying metastabillity, hysteresis and nonlinear dynamics [1]. In recent years the glass family has expanded to include systems that can be modelled by elastic manifolds in random potentials such as vortices in superconductors[14-20], domain walls [12] or two-dimensional electron layers[5,6]. When the random potential is weak these systems are expected to form a marginal glassy state, "Bragg glass", which is topologically ordered like a perfect crystal but unlike crystals, has no long-range spatial order[14,15]. An intriguing and enduring puzzle associated with this phase is the dynamics at the onset of motion: does it move as a rigid object or break up into pieces, does it crystallize at high velocities or retain its glassy nature[21-24].

To probe the dynamics, we focused on vortex states in single crystals of 2H-NbSe$_2$ because in this material quenched disorder can be sufficiently weak to allow the formation of a Bragg glass. The vortex states were prepared by field-cooling (FC) the sample below the superconducting transition in a field of 0.2 Tesla and temperatures down to 4.2 K. The results reported here were obtained on a sample of size 4.4x.0.8x0.006 mm$^3$ and transition temperature 7.2 K. At low temperatures (T<5.7K), where the Bragg glass is expected, the response of a freshly prepared FC lattice to a current pulse was previously[19] found to fit stretched exponential, or Kohlrausch-Williams-Watts (KWW) time dependence[10,11] spanning three decades in time: $V(t) = V_1 \left\{ 1 - \exp\left( -\left[ \frac{t-t_0}{\tau} \right]^\beta \right) \right\}$. Here $V_1$ is the saturation voltage, $t_0$ the delay time before a measurable voltage appears, $\tau$ the rise time and $\beta \sim 0.6$. The experimental protocol consists of applying a first current pulse (FP) of amplitude $I_1$ followed by a second pulse (SP) $I_2$, during which the evolution of the voltage is recorded (Fig1.b inset). The pulses are separated by a waiting time $t_w$ without current. Remarkably, the response during the SP is significantly *slower* than during the FP and its evolution depends not only on the elapsed time from the onset of $I_2$, as is the case in ergodic systems, but also on $t_w$, so $V(t) = V(t,t_w)$. This behaviour, also known as ageing, is one of the hallmarks of glassy dynamics[1-8]. The response curves for $I_2 = I_1$ and several values of $t_w$, are presented in Fig. 1a. When the same data is re-plotted against the scaled time $t/t_w$, (Fig. 1b), all the curves collapse without adjustable parameters onto a master curve,

$$V(t) = V_1 \left\{ 1 - \exp\left(-\left[\frac{t}{\gamma\, t_w}\right]^\beta\right) \right\} \qquad (1)$$

The scaling constant, $\gamma$, is independent of $t_w$ leading to a special and rare form of ageing $V(t,t_w) = V(t/t_w)$, also known as *simple* or *full* ageing [6-8]. Simple ageing is remarkably robust in this system extending over almost five decades in reduced time and holding up to the longest measurement times $\sim 2\, t_w$. For $T < 5.5$ K and at low saturation voltages, V $< 5\mu$V, the exponent $\beta$ is independent of $t_w$ and temperature. Its value, $\beta \sim 0.24$, obtained for $V_1 = 1.0$ μV, decreases slightly with increasing $V_1$ (Fig 1c). Simple ageing continues to hold at higher drives, but the range of the KWW fit is reduced. We find that the KWW function fits the data over a wider range than other simple choices. For example a logarithmic fit, also commonly used[6], is indistinguishable from KWW for t <0.1 $t_w$, but becomes worse at longer times. We note that for the SP, $\beta \leq 0.24$, is significantly lower than in the FP case, where $\beta \sim 0.6$. As shown below, this provides an important clue to the glassy dynamics of moving vortex states.

To study the case $I_1 \neq I_2$, $I_1$ was varied while keeping $I_2$ constant. The response is a sensitive function of $I_1$: it is slowest for $I_1 = I_2$ and becomes faster whenever the two levels are not equal (Fig. 2a). In other words, the system retains an imprint of $I_1$, which can be retrieved later in the form of a maximal slow-down during $I_2$. For t > 0.1 s, the response during $I_2$ fits a generalized form of Eqn. 1:

$$V(t) = V_2 \left\{ 1 - f \cdot \exp\left(-\left[\frac{t}{\gamma\, t_w}\right]^\beta\right) \right\} \quad (2)$$

Here $V_2$ is the SP saturation voltage and $f = f(V_1/V_2)$ is a "memory function". As shown in Fig. 2b, for each $V_2$, there exists a unique value $f$ that collapses the response for all $V_1$ onto one curve. The memory function $f(V_1/V_2)$ obtained by this procedure is plotted in Fig. 2c. Although the asymptotic response ($t > 0.1$ s) obeys equation 2, this form is not valid at short times (Fig. 2d).

Another limit, $t_w = 0$, was studied by applying "step-pulses" where the FP pulse $I_1$ was directly switched after a time $t_1$ to the SP pulse $I_2$. If the response is not allowed to saturate during the FP, the SP response is identical to that of a single pulse of amplitude $I_2$ with a shifted time origin : $V[t- (t_1 - \delta t)]$. The shift $\delta t$ is linear in $t_1$ (Fig. 3a inset), a behaviour that provides an additional clue to the mechanism underlying the glassy dynamics in this system.

KWW relaxation is far more common than the "conventional" exponential form ($\beta = 1$). It occurs in complex systems where the dynamics is governed by a statistical distribution of relaxation times together with constraints that restrict the path towards steady state to a hierarchical sequence of steps [9-11]. The hierarchy arises if certain segments (here chunks of vortices) cannot start moving until the ones in front of them are dislodged. Palmer *et al* [9] proposed a model of hierarchically constrained dynamics that leads to KWW response with $\beta = 1/(1 + \mu_0 \log 2)$ where $\mu_0$ is the number of degrees of freedom involved in initiating the process of relaxation. Thus different values of $\beta$ imply qualitative differences in the initial conditions, with smaller $\beta$ corresponding to more entangled states that require more steps to reach steady state. The exponents, $\beta \sim 0.6$ and $\beta \sim 0.2$, imply that the corresponding initial states for the FP and SP are inherently different. For the former, $\mu_0 \sim 1$, implies that the initial FC state is readily set in motion, while for the latter, $\mu_0 \sim 10$, indicates that the moving state (the SP is applied after the system experienced motion) is more entangled. This striking difference, together with the fact that the initial value, $\beta \sim 0.6$, cannot be recovered without warming up the sample, suggests structure of the FC state is altered irreversibly after the onset of motion. We propose that this is due to the introduction of dislocations when, due to pinning-potential inhomogeneities, some chunks of vortices start moving before others. As was shown in numerical simulations of driven 2D interacting systems[25], the dislocations minimize their energy by forming grain boundaries and by aligning their Burges vectors along the direction of motion. When the drive is suddenly removed they drift to restore the original state. But if annealing time scales are much longer than experimental times, the grain boundaries coarsen forming a more entangled spaghetti-like network of dislocations, resulting in a lower value of $\beta$.

It is generally accepted that the energy landscape of a finite disordered system has many local minima corresponding to metastable configurations surrounded by high energy barriers that can trap the system[8]. The trapping time in a metastable state increases with trap depth. In this context the dynamics of the vortex system can be modelled by mapping each state onto a point in configuration space and representing the evolution between two states by a connecting trajectory consisting of a sequence of trapped states. Thus, during the first pulse the system evolves from the FC state to the moving vortex state (MS) along a connecting trajectory as shown in Fig. 4. During $t_w$ when the drive is absent, the system drifts away from the MS point towards a lower energy Relaxed State (RS) where the grain boundaries have coarsened. Both simple ageing and the response to a step-pulse can be described within this model.

The key point for simple ageing is that the deepest traps encountered during $t_w$ must have trapping times $\tau_t \sim t_w$. This was shown to be the case[8] for trapping times that have exponential or power law distributions, provided the maximum trapping time is much shorter than $t_w$. Therefore during the subsequent SP, while the system is driven back towards the MS and traversing the same deepest trap, the trapping time should again be $\sim t_w$, provided the drive does not significantly change the energy landscape. In other words, $t_w$ selects a time scale (out of a broad distribution) which becomes the characteristic scale for subsequent response events. This naturally gives rise to simple ageing. However, in spite of its "simplicity" simple ageing is rare and was only recently observed in a Coulomb glass[5,6] and in a spin glass[7]. It is noteworthy that ageing may disappear altogether if the distribution of $\tau_t$ is not continuous or if it is truncated. For example in very disordered samples where $\tau_t >> t_w$, the system remains trapped close to the MS long after the drive is removed. This is the case for vortex states in Fe doped 2H-NbSe$_2$ where no ageing was observed for $t_w \leq 24$ hours[17]. At the other extreme lies the case of clean samples where ageing is not seen either because there is a unique equilibrium state (no trapping) or because $\tau_t$ is much shorter than the measurement times. This implies that there is a critical amount of disorder needed to observe ageing (see Supplementary Information).

The response to step-pulses imposes two additional constraints: (a) For a given FP amplitude $I$, the configuration space "speed", $v(I)$, along the *FC- MS* trajectory is constant. (b) $v(I)$ increases with increasing $I$. Thus during the FP the system evolves at an average speed $v_1 = v(I_1)$ so that at time $t_1$ it reaches an intermediate point $P$ along *FC-MS*. During the SP the remainder trajectory is traversed at a higher speed $v_2 = v(I_2)$. Had the entire *FC-MS* trajectory been traversed at speed $v_2$, then P would have been reached at a time $\delta t = t_1 (v_1/v_2)$ after the pulse onset. Therefore the response during SP, $V'(t- (t_1 - \delta t))$, is identical to that for a single pulse of amplitude $I_2$ applied $\delta t$ prior to $t_1$.

The experiments described here demonstrate that in the presence of quenched disorder the response of a driven vortex system to a current pulse can be described by

KWW time dependence, with the exponent reflecting the deviation of the initial state from equilibrium. It is shown that there exists a range of strengths of the quenched disorder for which the system can exhibit ageing and that simple ageing arises naturally in samples with a continuous distribution of trapping times whose range is much wider than that of experimental waiting times.

Correspondence and request for materials should be addressed to E.Y.A. (eandrei@physics.rutgers.edu).

## ACKNOWLEDGEMENTS

We wish to thank E. Abrahams, T. Giamarchi, E. Lebanon, M. Markus, C. Olson Reichhardt, A. Rosch and B. Rosenstein for useful discussions. Work supported by NSF-DMR-0456473 and by DOE DE-FG02-99ER45742.

## LIST OF REFERENCES

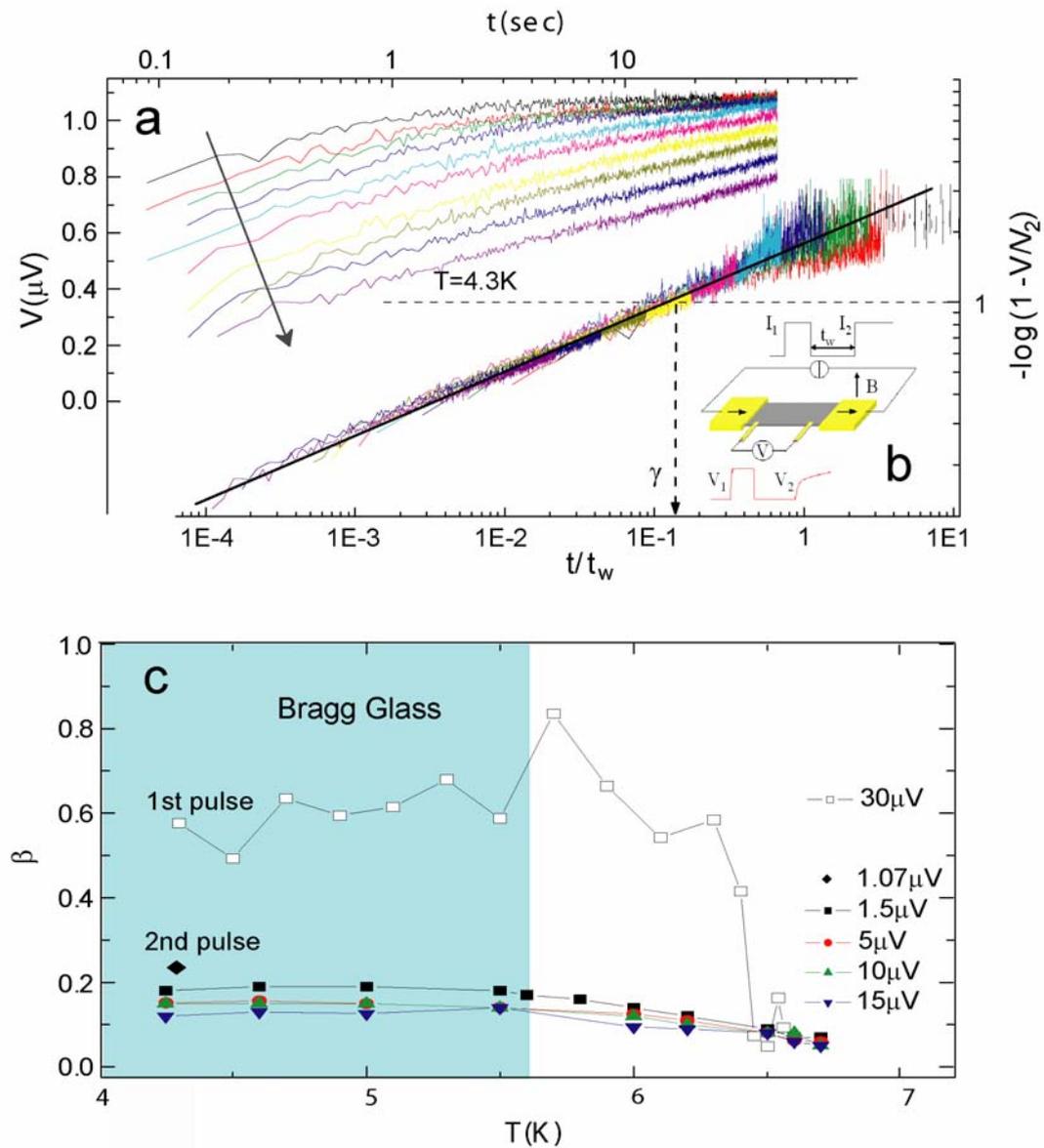

**Fig. 1**. **a**, **Ageing of the vortex lattice.** Response during the Second Pulse (SP) following a First Pulse (FP) of duration $t_1$ = 512 s and amplitude $I_1 = I_2$ = 5.36 mA. The waiting times $t_w$ = 4s, 8s, 16s, 32s, 64s, 128s, 256s, 512s, 1024s, 2048s increase along the arrow. **b**, Scaled SP response versus scaled time. A linear fit gives $\beta$ (slope) and of $\gamma$ from the intercept, $-log(1-V/V_1) = 1$. The experimental setup is shown in the inset. **c,** Temperature dependence of $\beta$ for FP and SP (open and solid symbols). Pulse amplitudes were adjusted to give the same saturation voltage at all temperatures.

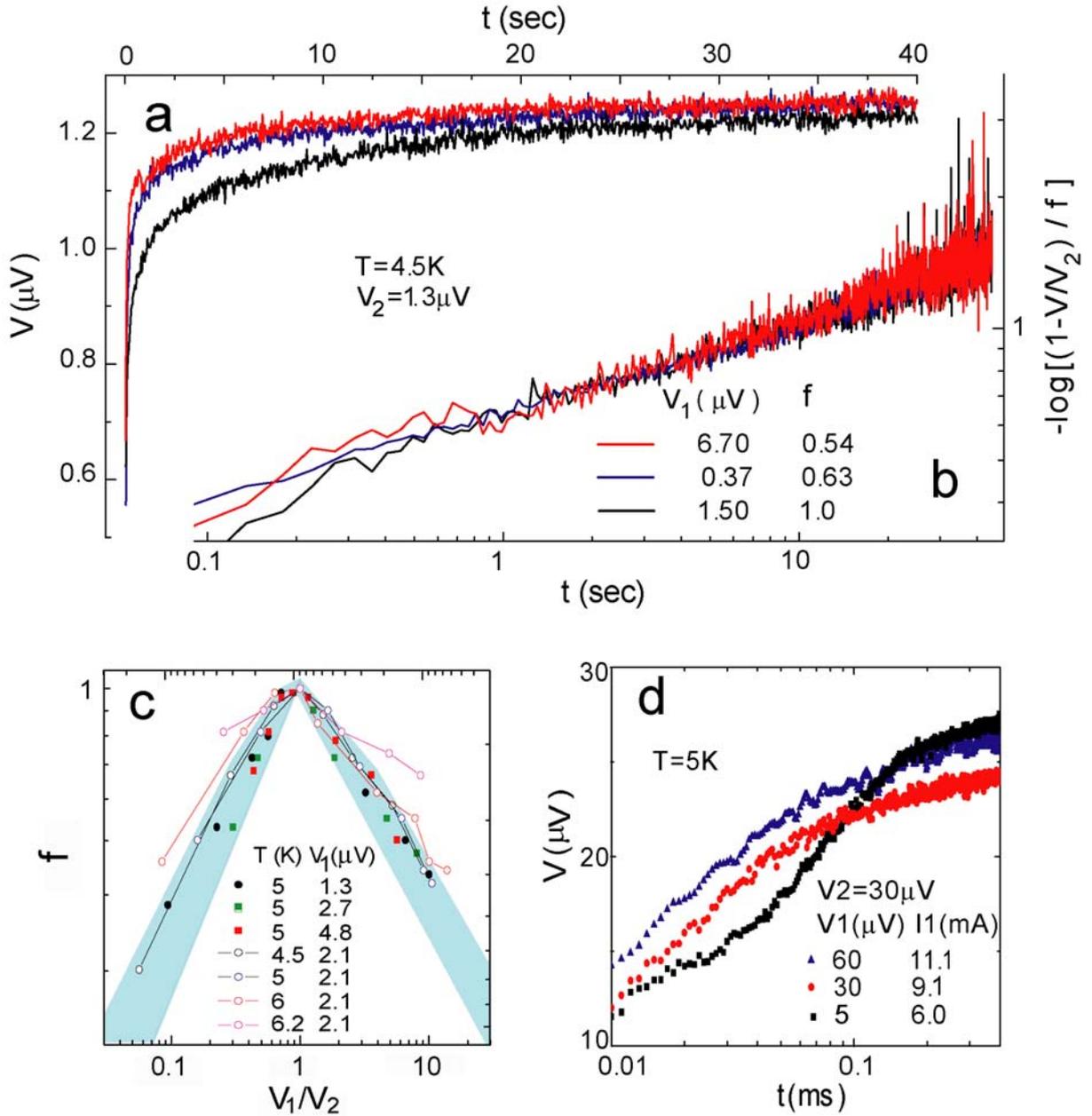

**Fig. 2. a**, **Memory of Pulse amplitude.** Time evolution of Second Pulse (SP) for $V_2 =$ 1.3 μV. **b,** Same data as in (a) showing that there exists a value, $f(V_1 / V_2)$, for which the scaled data, $-log[(1-V/V_2)/f]$ collapse onto a master curve. **c,** The memory function, $f(V_1 / V_2)$ obtained as described in (b). Highlighted area encloses data taken in the Bragg

glass regime, where memory is strongest.. For T > 5.7K, $f$ flattens out signalling a more feeble memory. **d**, Response in the first 0.3 ms of the SP, for $t_w = 240$ s, $I_2 = 9.1$ mA showing strong FP memory.

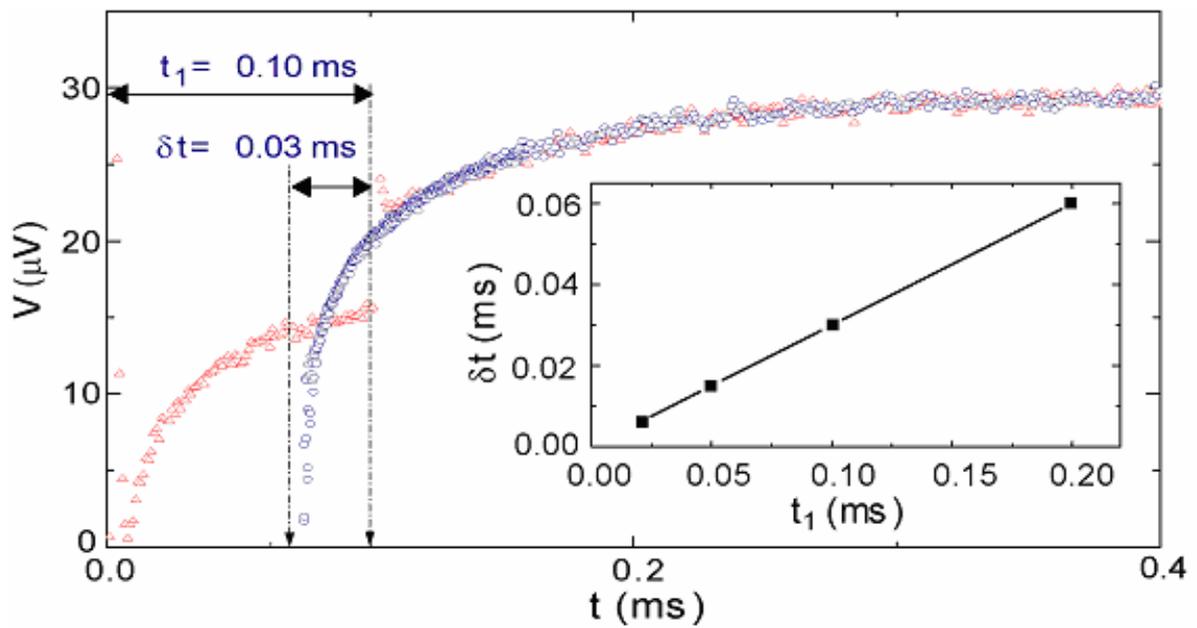

**Fig.3. Step-Pulse Response.** Response to step-pulse ($I_1 = 8.12$ mA $I_2 = 9.13$ mA $t_w=0$). The Second Pulse (SP) response (triangles) is compared to the response to a single pulse with the same current level $I_2$ (circles). The two curves overlap when shifting the time axis by $t_1-\delta t$. The inset shows that $\delta t$ is linear in $t_1$.

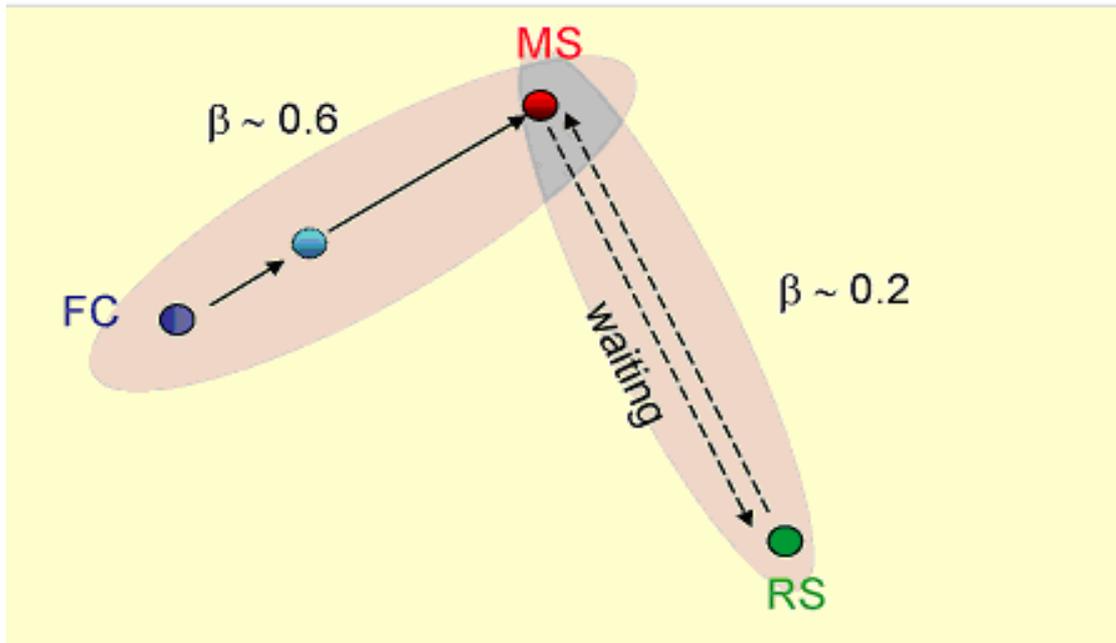

**Fig.4. Trajectories in configuration space.** Configuration space representation of vortex states and connecting trajectories. During the first pulse the system evolves along the FC – MS (Field Cooled - Moving State) trajectory which is independent of driving force. In between pulses the system drifts toward the Relaxed State (RS). During the SP, it is driven back to the MS.

# *Supplementary Information*

## *A. Sample details*

Samples used in the measurements were cleaved from an undoped single crystal of 2H-NbSe$_2$ and then into rectangular bars with dimensions 4.4 mm × 0.8 mm × 6 μm. The transition temperature and transition width of the samples are Tc = 7.2 K and ΔTc = 130 mK, respectively. The magnetic field was applied along the c-axis, and the transport current was applied in the a-b plane.

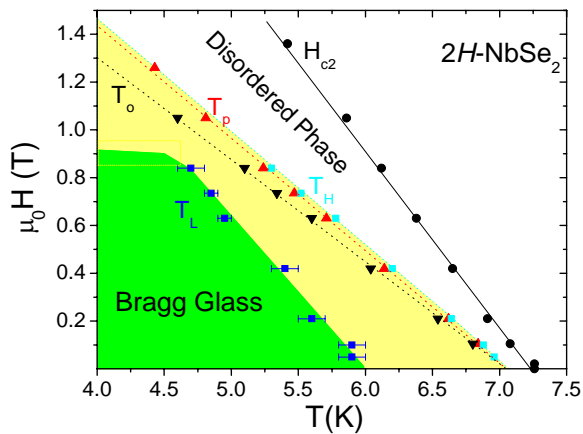

Supplementary Figure 1: **Phase diagram of the 2*H*-NbSe$_2$ sample (Fig.4 in PRL96, 017009 (2006)).**

Figure 1 shows the phase diagram obtained from the samples used in the glassy dynamics study. In this letter, we focus on the Bragg glass region (green area).

## *B. Measurement Setup*

The measurement setup was adapted to the signal level and rise time. For low saturation voltages (<5μV), where the voltage response can normally be measured with ~100ms time resolution, we used a Keithley 2400 source-meter to apply current and a Keithley 2182 nanovoltmeter to measure voltage. The time resolution of the low-level measurement system was set to ~ 45 ms.

For higher saturation voltages (>10μV) where the response time is shorter (<10ms), the current pulses were generated by a BNC 625A arbitrary function generator. To measure the voltage response, we combined a low noise voltage amplifier (gain=1000, input noise: 4nV/Hz$^{1/2}$) with a Tektronix 2232 oscilloscope. Each set of the measurements were typically taken 16~64 times and the data were averaged for better signal to noise ratio.

*C. Preparation of vortex states*

In the FC procedure the sample was first warmed to 10 K (above Tc) in a fixed magnetic field generated by a superconducting magnet working in persistent mode, and then quenched to the measurement temperature T. The duration of a typical FC cycle was ~ 5 minutes. The pulse measurement was carried out about 60s after the measurement temperature was reached.